\documentclass[conference]{IEEEtran}
\IEEEoverridecommandlockouts
\usepackage{cite}
\usepackage{amsmath,amssymb,amsfonts}
\usepackage{algorithmic}
\usepackage{graphicx}
\usepackage{booktabs, tabularx}
\newcolumntype{C}{>{\centering\arraybackslash}X}
\usepackage{textcomp}
\usepackage{xcolor}

\def\cN{{\cal N}}

\def\C{{\mathbb C}}

\DeclareMathOperator*{\argmax}{argmax\,}
\DeclareMathOperator*{\argmin}{argmin\,}
\newcommand\norm[1]{\lVert#1\rVert}
\newcommand\nth[1]{$#1^{\mathrm{th}}$}

\def\BibTeX{{\rm B\kern-.05em{\sc i\kern-.025em b}\kern-.08em
    T\kern-.1667em\lower.7ex\hbox{E}\kern-.125emX}}
\begin{document}

\title{MRI Reconstruction with Side Information using Diffusion Models\\

\thanks{This work was supported by NSF IFML 2019844, NSF CCF-2239687 (CAREER), NIH U24EB029240, ARO W911NF2110117, and Oracle for Research Fellowship.}
}

\author{\IEEEauthorblockN{1\textsuperscript{st} Brett Levac}
\IEEEauthorblockA{\textit{ECE} \\
\textit{University of Texas}\\
Austin, USA \\
blevac@utexas.edu}
\and
\IEEEauthorblockN{2\textsuperscript{nd} Ajil Jalal}
\IEEEauthorblockA{\textit{EECS} \\
\textit{University of California}\\
Berkeley, USA \\
ajiljalal@berkeley.edu}
\and
\IEEEauthorblockN{3\textsuperscript{rd} Kannan Ramchandran}
\IEEEauthorblockA{\textit{EECS} \\
\textit{University of California}\\
Berkeley, USA \\
kannanr@eecs.berkeley.edu}
\and
\IEEEauthorblockN{4\textsuperscript{th} Jonathan I. Tamir}
\IEEEauthorblockA{\textit{ECE} \\
\textit{University of Texas}\\
Austin, USA \\
jtamir@utexas.edu}
% \and
% \and
}

\maketitle

\begin{abstract}
Magnetic resonance imaging (MRI) exam protocols consist of multiple contrast-weighted images of the same anatomy to emphasize different tissue properties. Due to the long acquisition times required to collect fully sampled k-space measurements, it is common to only collect a fraction of k-space for each scan and subsequently solve independent inverse problems for each image contrast. Recently, there has been a push to further accelerate MRI exams using data-driven priors, and generative models in particular, to regularize the ill-posed inverse problem of image reconstruction. These methods have shown promising improvements over classical methods. However, many of the approaches neglect the additional information present in a clinical MRI exam like the multi-contrast nature of the data and treat each scan as an independent reconstruction. In this work we show that by learning a joint Bayesian prior over multi-contrast data with a score-based generative model we are able to leverage the underlying structure between random variables related to a given imaging problem. This leads to an improvement in image reconstruction fidelity over generative models that rely only on a marginal prior over the image contrast of interest. 
\end{abstract}


\section{Introduction}
\label{sec:intro}
 Magnetic Resonance Imaging (MRI) is a medical imaging technique that has superior soft tissue contrast without the use of ionizing radiation. Unfortunately, MRI acquisition times are slow due to hardware constraints and resolution requirements. To reduce scan times, a common approach is to collect less data and subsequently solve an ill-posed inverse problem for reconstruction. A variety of advancements have been developed to address the inverse problem such as parallel imaging
 \cite{sense},
 % \cite{smash, sense, grappa},
 compressed sensing \cite{lustig2007sparse},
 % low-rank modeling \cite{LORAKS,trzasko}, 
 and more recently data-driven methods 
 \cite{hammernik, modl, jalal2021robust, CHUNG_score, martin_score, Bendel_2022}.
 % \cite{hammernik, modl, Sriram2020EndtoEndVN, deepjsense, jalal2021robust, CHUNG_score, martin_score, Bendel_2022}.
 Data driven techniques have shown marked improvements over prior approaches and are closing in on clinical validation 
 \cite{J_Prosp_2023}.
 % \cite{recht2020using,J_Prosp_2023}.

In MRI, different image weightings arise due to a variety of underlying tissue properties. These contrasts can be observed by modifying pulse sequence parameters to increase sensitivity to a desired tissue property. Common image contrasts include Proton Density (PD), $T_1$, and $T_2$ weightings, which are primarily based on tissue density and relaxation. Additional  contrasts like diffusion weighted imaging (DWI) create images sensitive to water diffusion through tissue which can be useful in tumor identification \cite{DWI}; susceptibility weighted imaging (SWI) creates contrast based on magnetic susceptibility changes in the body that can often occur in the brain due to microbleeds \cite{SWI}. Each of these contrasts can assist in diagnosing and assessing different underlying health conditions.

\begin{figure}[h!]
\begin{center}
  \includegraphics[width=.8\columnwidth]{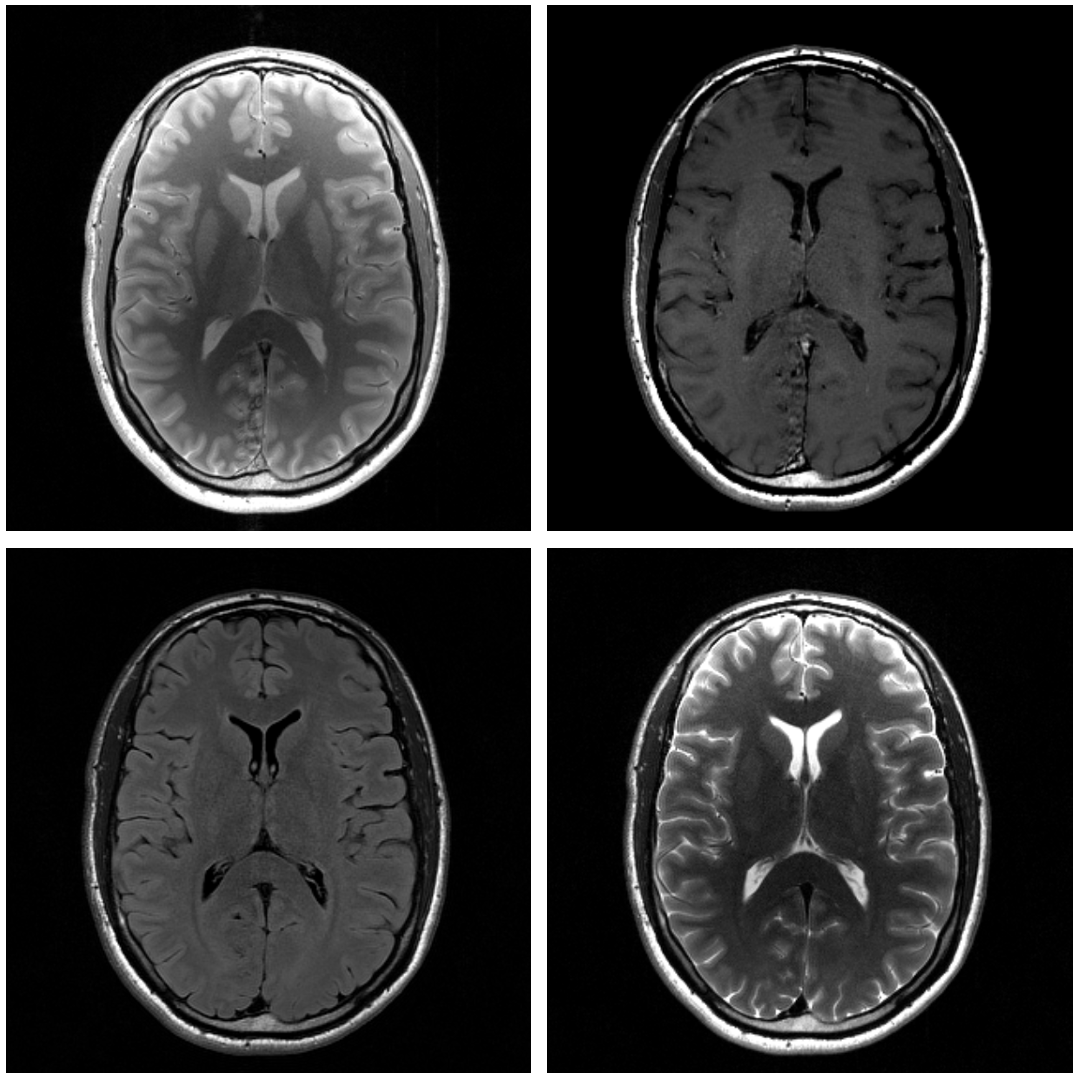}
\end{center}
\caption{Example of a single brain slice with four different contrasts. Proton Density (Top Left), T1 (Top Right), T2-Flair(Bottom Left), T2 (Bottom Right).}
\label{fig:contrast}
\end{figure}

In a typical MRI exam, data representing different contrast-weighted images are often serially collected, as illustrated in Fig. \ref{fig:contrast}. Image reconstruction is typically done independently for each contrast. However, there is clearly shared information between image contrasts for the same anatomy which can be exploited to assist the reconstruction of one, or all, of the images. Past approaches have leveraged this basic property through handcrafted priors 
% \cite{Bilgic2011,Ehrhardt2016,HuangJunzhou2014FmMr}
\cite{Bilgic2011,Ehrhardt2016}, end-to-end deep learning %\cite{DuDorNet,Lei2019,LIU2021,Do2020,jVN, SunLiyan2019ADIS,Do2020}
\cite{DuDorNet,Lei2019,jVN}, and generative models such as GANs 
% \cite{Salman2020 ,kelkar2021prior,Kelkar2022,bitingyu,hinet,generative_review}
\cite{Salman2020 ,kelkar2021prior} to reduce data acquisition and enhance multi-contrast image reconstruction. Methods which utilize handcrafted priors typically rely on regularization terms based on feature similarity in a transform domain. Meanwhile, end-to-end deep learning techniques attempt to couple the reconstruction processes of different images through shared reconstruction weights. Finally, GAN based approaches can range from learning a synthesis function mapping one image contrast to another or even modified Compressed Sensing using Generative Modeling (CSGM) frameworks.
% \cite{kelkar2021prior,Kelkar2022,Salman2020} 

In this work we derive a joint Bayesian prior over multi-contrast data that is learned using score-based generative models \cite{song_ermon} and can be used to solve two closely related problems: \textbf{(1)} Joint reconstructions, where multiple image contrasts are under-sampled; and \textbf{(2)} Conditional reconstructions, where one or more of the image contrasts are fully sampled. We specialize our approach to the case where two image contrasts are scanned and show that learning joint priors over contrast pairs can enhance reconstruction for both joint reconstruction and prior image constrained reconstruction without retraining for new sampling trajectories or acceleration levels.

\section{Background}

% \subsection{Problem Formulation}
% We assume an MRI exam consists of measuring $n$ of the same anatomical images each with different contrast weighting. The measurements for the \nth{i} contrast is described by
% \begin{equation}
% \label{eqn:forw1}
%     y_i = P_i FSx_i + \epsilon_i,
% \end{equation}
% where $S$ denotes coil sensitivity maps, $F$ is the Fourier transform, $P_i$  are independent sampling operators for each contrast, $x_i$ is the \nth{i} contrast image,  and $\epsilon_i \sim \cN(0,\sigma^2 I_N)$ is an additive noise term. For the remainder of this work we will, without loss of generality, consider the case of $n=2$ contrast images and refer to the forward operators as $A_1=P_1FS$ and $A_2=P_2FS$.
% Under this setting, we consider two different problems: \textbf{(1)} Recover $x_1$ and $x_2$ when both sets of measurements ($y_1,y_2$) are under-sampled, \textbf{(2)} Recover $x_1$ when $y_1$ is under-sampled and $y_2$ is fully sampled giving us access to the true image $x_2$.

\subsection{Problem Formulation}
For a desired vectorized image $x \in \C^N$ the corresponding MRI measurements $y \in \C^M$ are acquired in the Fourier domain (k-space). Letting $\mathcal{F}\in\C^{M\times N}$ denote a Fourier encoding matrix (2D, 3D, Cartesian, or non-Cartesian), the measured data $y$ in the single-coil imaging case is given by
\begin{equation}
y = \mathcal{F}x + \epsilon,
\end{equation}
where $\epsilon \sim \cN (0,\sigma^2 I_M)$ is zero-mean complex-valued white Gaussian noise. In the case of Cartesian sampling, we can write $\mathcal{F}=PF$ where $F\in\C^{N\times N}$ is the multi-dimensional DFT matrix, and $P \in \{0,1\}^{M\times N}$ is a binary selection (diagonal) matrix.

Multi-contrast imaging entails measuring $n_c$ images of the same anatomical structures each with different contrast weightings. In principle, these contrasts will also vary according to imaging orientation (sagittal, axial, coronal). The measurements for the \nth{i} contrast are described by
\begin{equation}
\label{eqn:forw1}
    y_i = \mathcal{F}_ix_i + \epsilon_i,
\end{equation}
where $\mathcal{F}_i$  are independent Fourier encoding operators (potentially different sampling schemes) for each contrast, $x_i$ is the \nth{i} contrast image,  and $\epsilon_i \sim \cN(0,\sigma_i^2 I_M)$ is an additive noise term independent between contrasts. For the remainder of this work, we will without loss of generality consider the case of two aligned contrast images ($n_c = 2$) with $\sigma_1=\sigma_2=\sigma$  and refer to the forward operators as $\mathcal{F}_1$ and $\mathcal{F}_2$.
Under this setting, we consider two related problems: \textbf{(1)} Recover $x_1$ and $x_2$ when both sets of measurements ($y_1,y_2$) are under-sampled; \textbf{(2)} Recover $x_2$ when $y_2$ is under-sampled and $y_1$ is fully sampled giving us access to $x_1$ from a conventional reconstruction.

\subsection{Optimization Approaches}
Several prior works have addressed the challenges of joint and conditional reconstructions from multi-contrast data. Before the use of deep learning in medical image reconstruction, most techniques leveraged hand crafted regularization terms to couple the reconstruction process for multiple images with the assumption that there are shared structural details between images (e.g., edges, image gradients, low-rank representation)
%\cite{Bilgic2011,Ehrhardt2016,HuangJunzhou2014FmMr,PROMISE, ref_MRI, Bustin_LR, Hankel_joint, Berkin_2018}
\cite{Bilgic2011, Ehrhardt2016,Bustin_LR}. Ultimately these methods set out to solve an optimization problem of the following form:
% \begin{equation}
% \label{eqn:conv_prior_CS}
% x_1^*,x_2^* = \argmin_{x_1,x_2} \norm{\mathcal{F}_1x_1-y_1}^2_2 + \norm{\mathcal{F}_2x_2-y_2}^2_2 + \lambda R(x_1,x_2),
% \end{equation}

\begin{multline}
\label{eqn:conv_prior_CS}
x_1^*,x_2^* = \argmin_{x_1,x_2} \norm{\mathcal{F}_1x_1-y_1}^2_2 + \norm{\mathcal{F}_2x_2-y_2}^2_2 \\ + \lambda R(x_1,x_2),
\end{multline}

where $R(x_1, x_2)$ is a regularization functional that couples the image contrasts and $\lambda>0$ is the regularization weighting.
A subset of prior works justify Eq.~\ref{eqn:conv_prior_CS} under a Bayesian framework with the goal of solving maximum a posteriori (MAP) estimation \cite{Bilgic2011, Ehrhardt_2015}. Under the forward model in Eq.~\ref{eqn:forw1}, joint image reconstruction via MAP takes the form:
% \begin{align}
% \label{eqn:MAP_joint}
% x_1^*, x_2^* &= \argmax_{x_1,x_2} \log{p(x_1,x_2|y_1,y_2)},\\ 
% &= \argmax_{x_1,x_2} \log{p(y_1,y_2|x_1,x_2)} + \log{p(x_1,x_2)},\\
% &= \argmin_{x_1,x_2} \norm{\mathcal{F}_1x_1-y_1}_2^2 + \norm{\mathcal{F}_2x_2-y_2}_2^2 -  \log{p(x_1,x_2)}.
% \end{align}
\begin{align}
% \label{eqn:MAP_joint}
x_1^*, x_2^* &= \argmax_{x_1,x_2} \log{p(x_1,x_2|y_1,y_2)}\\ 
&= \argmax_{x_1,x_2} \log{p(y_1,y_2|x_1,x_2)} + \log{p(x_1,x_2)}.
\end{align}
upon including the log-likelihood model for each measurement process we obtain
\begin{multline}
\label{eqn:MAP_joint}
x_1^*, x_2^* = \argmin_{x_1,x_2} \norm{\mathcal{F}_1x_1-y_1}_2^2 + \norm{\mathcal{F}_2x_2-y_2}_2^2 \\ -  \log{p(x_1,x_2)}.
\end{multline}

The hand-crafted priors used are approximations to the underlying statistical prior and are therefore limited in their performance. It has also been shown that solving for MAP, even with more accurate priors, can lead to biased solutions \cite{Ajil_2021_ICML}. 
Approaches which admit optimization problems different than Eq. \ref{eqn:conv_prior_CS} include dictionary learning \cite{joint_dict} and physics constrained reconstructions \cite{T2Shuff_spm}. Joint dictionary learning builds upon single image dictionary learning \cite{single_dict} by creating a coupled dictionary between multi-contrast images. Knowledge about signal dynamics can also be used to constrain the multi-contrast optimization problem. This is done by creating a set of temporal basis functions from a dictionary of signal decay curves and subsequently estimating basis coefficients rather than individual multi-contrast images \cite{T2Shuff_MRM,Christodoulou2018}.

\subsection{Deep Learning Approaches}
More recently, deep learning approaches for MRI reconstruction have been extended to  handle multi-contrast data. Many of these methods use end-to-end training with shared weights between multi-contrast data channels to enhance the sharing of information during either prior image constrained reconstruction
% \cite{DuDorNet,Lei2019,LIU2021,Do2020,xiang2018ultra}
\cite{DuDorNet,Lei2019}
or joint reconstruction 
% \cite{jVN, SunLiyan2019ADIS,Do2020}
\cite{jVN}. 
% Self-supervised learning with transformer architectures has also been used to deal with  partially sampled data, along with shared weights for multi-contrast reconstruction \cite{self_supervised_transformer}.

Another class of deep learning approaches have employed the use of generative models. These methods can be separated into \textbf{(1)} prior-image constrained reconstruction and \textbf{(2)} prior-image constrained synthesis. Methods for \textbf{(1)} are heavily inspired by the Compressed Sensing using Generative Models (CSGM) framework proposed for more general inverse problems \cite{CSGM-bora17}. One such approach attempts to reconstruct high fidelity target images from low resolution target data and high resolution reference images of an alternative image contrast \cite{Salman2020}. Later works leverage a StyleGAN architecture 
% \cite{StyleGAN,StyleGAN2}
to reconstruct from under-sampled target data and fully sampled reference images 
% \cite{kelkar2021prior,Kelkar2022}
\cite{kelkar2021prior}. Techniques addressing \textbf{(2)} aim to create a target image with access only to a reference image using conditional GANs~
% \cite{Nie_2018,Dar_2019,Armanious_2020,DiamondGAN_2019}
\cite{Dar_2019,DiamondGAN_2019}, CycleGANs~\cite{cyclemedgan}, or diffusion models~\cite{ozbey_2022}.  Our proposed method is of type \textbf{(1)} but can also be used for purely synthesis problems. 
\begin{figure}[!ht]
\begin{center}
  \includegraphics[width=.8\columnwidth]{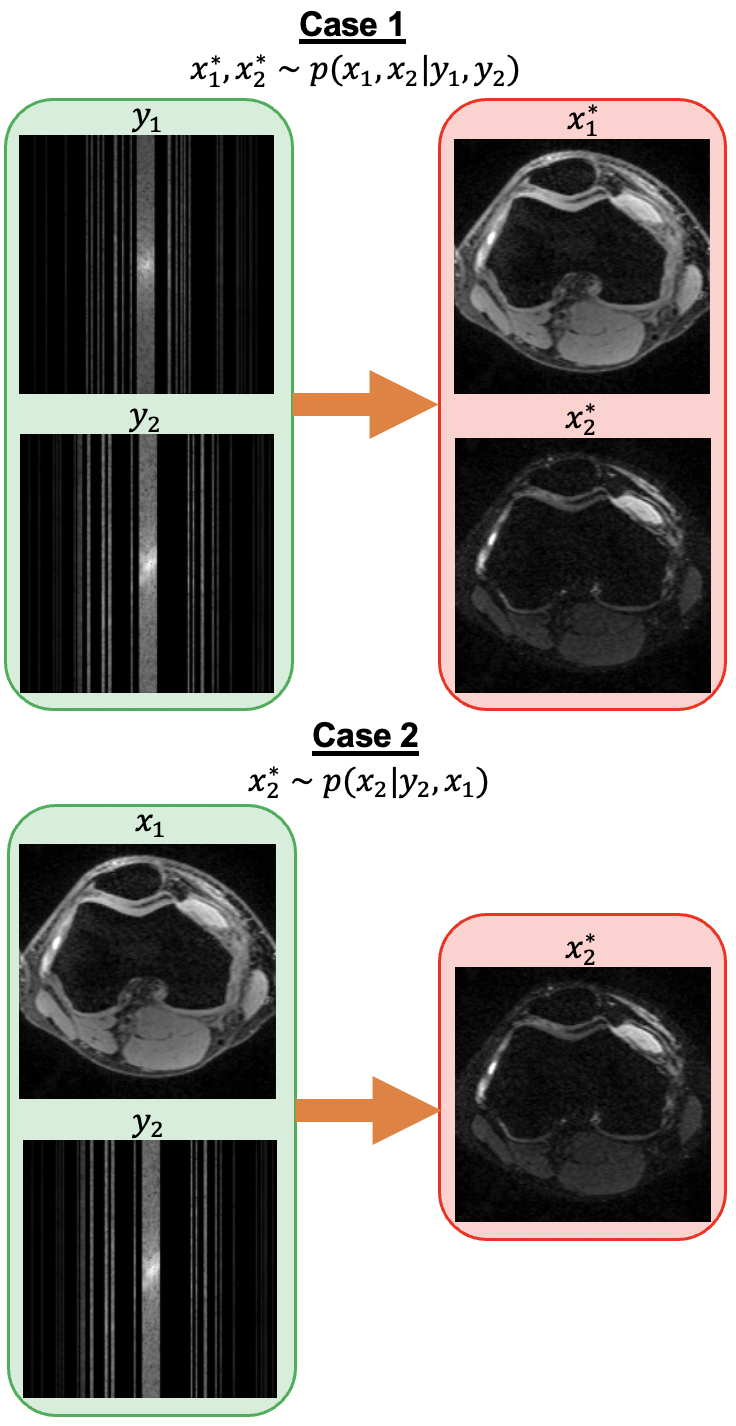}
\end{center}
\caption{Problem settings. (Top) Joint reconstruction with under-sampled measurements from both contrasts. (Bottom) Single image reconstruction with access to a fully sampled reference scan and under-sampled measurements from target scan.}
\label{fig:cases}
\end{figure}

\subsection{Score Models}
\label{subsubsec:score_models}
Recent work in deep learning has led to impressive progress in the area of generative modeling \cite{goodfellow2020generative,kingma2013auto}, and in particular, score-based generative models \cite{song_ermon}. These models aim to approximate the score of a distribution $p$, defined as $\nabla_x \log p(x)$, by training a neural network via denoising score matching with access only to  i.i.d training samples from the high-dimensional target distribution\cite{denoise_score, song_ermon}. These models produce new samples approximately from $p$ by initializing $x^0 \sim \cN(0,I_N)$, and following annealed Langevin dynamics for $T$ steps:
\begin{align}
    x^{t+1} = x^t + \eta^t \nabla_x^t \log p(x^t) + z^t,\label{eqn:langevin}
\end{align}
where $z^t \sim \cN(0,\,\sqrt{\eta^t} I_N)$ and $\eta^t$ is the variance of the additive noise at each step. As $\eta^t \to 0, T\to \infty$, the distribution of $x^T$ converges to $p$ \cite{song_ermon}.  
We can write similar updates for the posterior distribution, $p(x|y)$. If we know the likelihood distribution $y|x$, then using Bayes rule we can sample via:
\begin{equation}
\label{eqn:single_inf}
    x^{t+1} = x^t + \eta^t \nabla_{x^t} \left( \log p(y|x^t)+ \log p(x^t) \right) + z^t.
\end{equation}
These models have been used to solve inverse problems such as compressed sensing MRI\cite{jalal2021robust,CHUNG_score,martin_score, Gungor_2022}.

\section{Methods}
\label{sec:methods}

\subsection{Joint Reconstruction (Case 1)}
\label{subsec:joint_recon}
 Given under-sampled measurements for both images we can formulate our joint reconstruction algorithm as follows:
\begin{align}
    x_1^{t+1} =& x_1^t + \eta^t \nabla_{x_1^{t}} \log p(x_1^{t},x_2^t|y_1, y_2) + z_1^t, \\
    x_2^{t+1} =& x_2^t + \eta^t \nabla_{x_2^{t}} \log p(x_1^{t},x_2^t|y_1, y_2) + z_2^t.
\end{align}
Applying Bayes rule gives
% \begin{align}
%     x_1^{t+1} =& x_1^{t} + \eta^t  \nabla_{x_1^{t}} \left( \log p(y_1, y_2|x_1^{t},x_2^t) + \log p(x_1^t,x_2^t) \right) + z_1^t,\\
%     x_2^{t+1} =& x_2^{t} + \eta^t  \nabla_{x_2^{t}} \left( \log p(y_1, y_2|x_1^{t},x_2^t) + \log p(x_1^t,x_2^t) \right) + z_2^t.
% \end{align}

\begin{multline}
    x_1^{t+1} = x_1^{t} + \eta^t  \nabla_{x_1^{t}} \left( \log p(y_1, y_2|x_1^{t},x_2^t) + \log p(x_1^t,x_2^t) \right) \\ + z_1^t,
\end{multline}
\begin{multline}
    x_2^{t+1} = x_2^{t} + \eta^t  \nabla_{x_2^{t}} \left( \log p(y_1, y_2|x_1^{t},x_2^t) + \log p(x_1^t,x_2^t) \right) \\ + z_2^t.
\end{multline}

By substituting the approximate likelihood equations and the neural network score function approximator ($s_\theta(x_1,x_2)_i \approx \nabla_{x_i} \log{p(x_i,x_j)}$) we obtain 

\begin{align}
    x_1^{t+1} =& x_1^{t} + \eta^t   \left( \frac{\mathcal{F}_1^H(\mathcal{F}_1 x_1^t-y_1)}{\gamma_t^2 + \sigma^2} + s_\theta(x_1^t,x_2^t; \beta_t )_1 \right) + z_1^t,\\
    x_2^{t+1} =& x_2^{t} + \eta^t   \left( \frac{\mathcal{F}_2^H(\mathcal{F}_2 x_2^t-y_2)}{\gamma_t^2 +\sigma^2} + s_\theta(x_1^t,x_2^t; \beta_t )_2 \right) + z_2^t,
\label{eqn:joint_inf1}
\end{align}

\noindent where $\beta_t$ denotes the current noise level during inference and $\gamma_t$ is an annealing term which is included to try and account for the missmatch between the liklihood $\nabla_{x} \log{p(y|x)}$, which we know exactly,  and $\nabla_{x^t} \log{p(y|x^t)}$.

\subsection{Conditional Reconstruction (Case 2)}
\label{subsec:cond_recon}
 In Case 2 our goal is to conduct posterior sampling conditioned on under-sampled k-space data ($y_2$) and the reconstruction from a fully-sampled prior scan (i.e., $x_1$). When $\epsilon_1$ is small, we can approximate $x_1^*\approx \mathcal{F}_1^{\dagger} y_1$, where ${F}_1^{\dagger}$ is the pseudo-inverse of the forward.
 % In practice $x_1$ is derived from (potentially) noisy measurements however when we assume that when $\epsilon_2$ is small we can approximate $x_1$ as $\mathcal{F}^Hy_1$. 
 Written in the Langevin update equations, this task admits the following simplified algorithm:
\begin{equation}
    x_2^{t+1} = x_2^{t} + \eta^t \nabla_{x_2^{t}} \log p(x_2^{t}|y_2, x_1^*) + z_2^t.
\end{equation}
Using Bayes rule we can get the following equation:
\begin{equation}
    x_2^{t+1} = x_2^t + \eta^t \nabla_{x_2^t} \left( \log p(y_2,x_1^*|x_2^t) + \log p(x_2^t) \right) + z_2^t,
\end{equation}
 Looking at the measurement process for both images we see in Eq. \ref{eqn:forw1} that given $x_2$ the measurements $y_2$ are only stochastic due to the additive noise $\epsilon_2$. As noise between MRI scans is independent, $y_2$ is conditionally independent of $x_1$ given $x_2$.  This leaves
% \begin{align}
%     & x_2^{t+1} = x_2^t + \eta^t \nabla_{x_2^t} \left( \log p(y_2|x_2^t) +\log p(x_1^*|x_2^t)+ \log p(x_2^t) \right) + z_2^t,\\
%     \implies& x_2^{t+1} = x_2^t + \eta^t \nabla_{x_2^t} \left( \log p(y_2|x_2^t) + \log p(x_1^*,x_2^t) \right) + z_2^t.
% \end{align}

\begin{multline}
    x_2^{t+1} = x_2^t + \eta^t \nabla_{x_2^t} \left( \log p(y_2|x_2^t) +\log p(x_1^*|x_2^t) \right. \\
    \left. + \log p(x_2^t) \right) + z_2^t
\end{multline}
\begin{multline}
    x_2^{t+1} = x_2^t + \eta^t \nabla_{x_2^t} \left( \log p(y_2|x_2^t) + \log p(x_1^*,x_2^t) \right) + z_2^t.
\end{multline}

Using the analytically available derivative of the likelihood function we can further simplify the update equation as 

% \begin{align}
%     &x_2^{t+1} = x_2^t + \eta^t \left( \frac{\mathcal{F}_2^H(\mathcal{F}_2 x_2^t-y_2)}{\sigma^2}+\nabla_{x_2^t}\log p(x_1^*, x_2^t) \right)+ z^t,\\
%     \implies& x_2^{t+1} = x_2^t + \eta^t \left( \frac{\mathcal{F}_2^H(\mathcal{F}_2 x_2^t-y_2)}{\sigma^2}+s_\theta(x_1^* + \lambda_t, x_2^t; \beta_t )_2 \right)+ z^t,
%     \label{eqn:cond_inf}
% \end{align}

\begin{multline}
    x_2^{t+1} = x_2^t + \eta^t \left( \frac{\mathcal{F}_2^H(\mathcal{F}_2 x_2^t-y_2)}{\gamma_t^2 + \sigma^2} \right. \\
    \left.  \vphantom{\frac{\mathcal{F}_2^H(\mathcal{F}_2 x_2^t-y_2)}{\sigma^2}} +\nabla_{x_2^t}\log p(x_1^*, x_2^t) \right)+ z^t,
\end{multline}

\begin{multline}
    x_2^{t+1} = x_2^t + \eta^t \left( \frac{\mathcal{F}_2^H(\mathcal{F}_2 x_2^t-y_2)}{\gamma_t^2 + \sigma^2} \right. \\
    \left. \vphantom{\frac{\mathcal{F}_2^H(\mathcal{F}_2 x_2^t-y_2)}{\gamma_t^2 + \sigma^2}}+s_\theta(x_1^* + \lambda_t, x_2^t; \beta_t )_2 \right)+ z^t,
    \label{eqn:cond_inf}
\end{multline}

% \begin{equation}
%     \hat{x}_{t+1} = \hat{x}_t + \eta_t \left( \frac{\hat{A}^H(\hat{A}\hat{x}_t-y)}{\sigma^2}+s_\theta(\hat{x}_t, \tilde{x} + \lambda_t;\beta_t) \right)+ z_t,
% \end{equation}

\noindent where $\lambda_t \sim \cN(0,\beta_t^2 I_n)$ is an additive noise term dependent the noise level at the current inference step during the annealed sampling process. The difference between Eq. \ref{eqn:single_inf} and our new update equations (Eq. \ref{eqn:joint_inf1}, Eq. \ref{eqn:cond_inf}) is that we now need access to the joint score function $\nabla_{x_i^t}\log p(x_i^t, x_j^t)$.

\subsection{Experiments}
\label{sec:exp}
\begin{figure}[h!]
\begin{center}
  \includegraphics[width=.85\columnwidth]{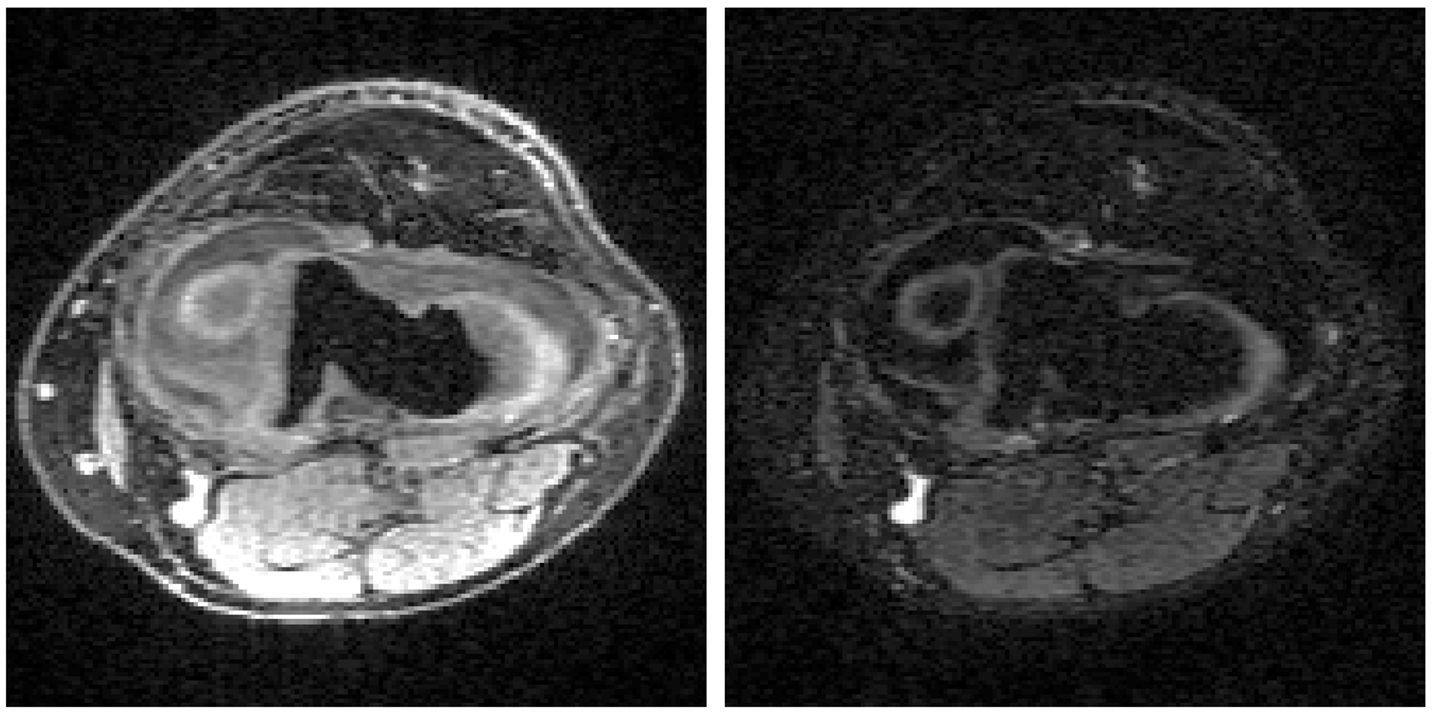}
\end{center}
\caption{Example of an image pair in the publicly available dataset used for training and testing \cite{skm-tea}. (Left) $x_1$, (Right) $x_2$. The dynamic range for both subfigures is the same.}
\label{fig:skm_ex}
\end{figure}

\begin{figure}[h!]
\begin{center}
  \includegraphics[width=.85\columnwidth]{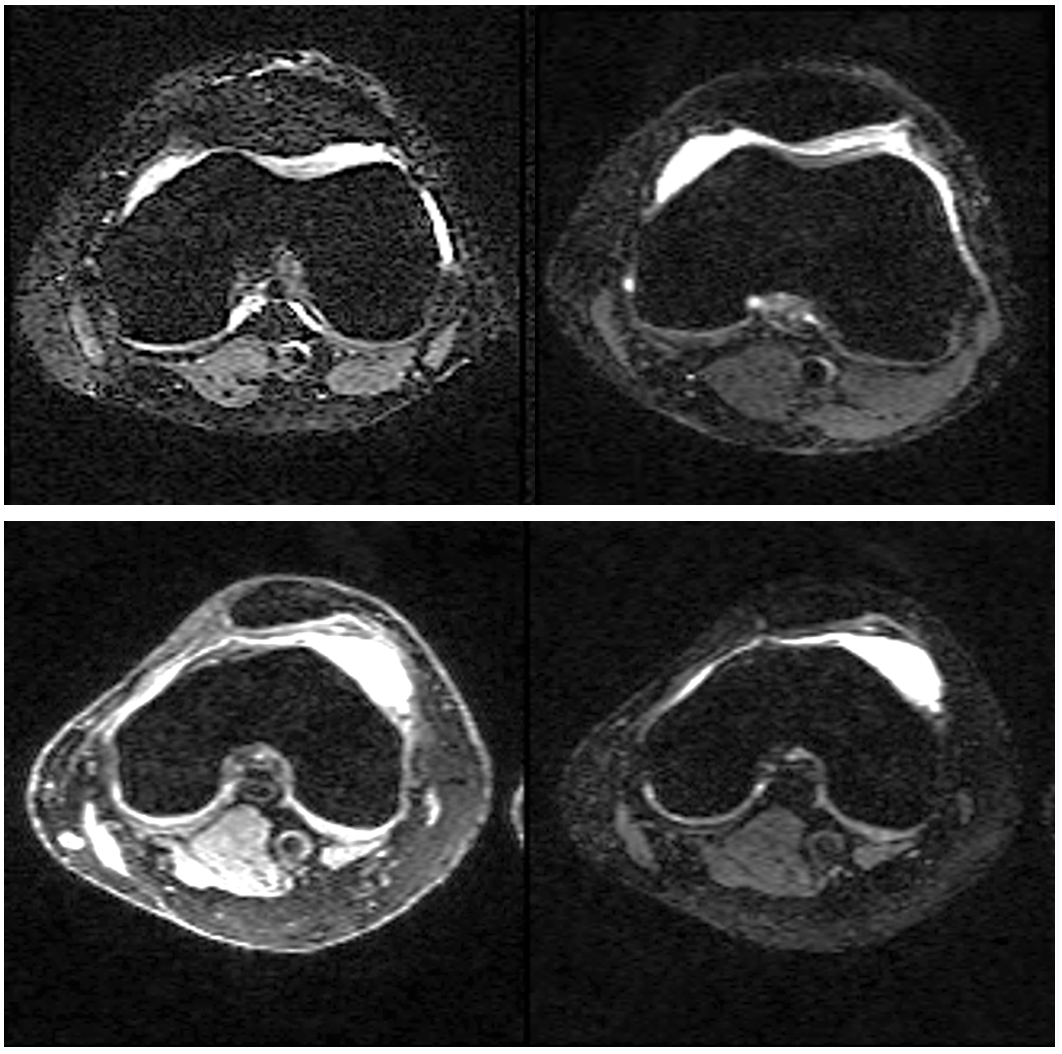}
\end{center}
\caption{Two prior samples from marginal score model (Top), One prior sample from joint score model (Bottom).}
\label{fig:prior_samples}
\end{figure}

For proof-of-principle, we used the publicly available SKM-TEA dataset \cite{skm-tea} which contains paired images from double-echo in steady state (DESS) gradient echo knee exams consisting of 3D multi-coil k-space data collected at two different echo times as well as reference complex-valued reconstructions based on parallel imaging. An example of paired images is show in Fig. \ref{fig:skm_ex}, noting the weaker intensity of the second echo due to signal relaxation. Training data pairs $(x_1,x_2)$ were normalized jointly by the \nth{99} percentile value of the reference magnitude image for $x_1$. We trained two different score models, $s_\theta(x_2)$ and $s_\theta(x_1,x_2)$, which model the marginal and joint score functions respectively. The input sizes to the marginal and joint score networks were $2$ and $4$ channels to account for real and imaginary components of one or two images respectively. We used $100$ slices from $150$ different volumes ($15,000$ slices in total) to train the networks. The networks were trained for $80,000$ steps with a batch size of $4$. Representative random samples taken via the annealed Langevin dynamics update equations (Eq. \ref{eqn:langevin}) \cite{song_ermon} are shown for both the marginal and joint score models in   Fig. \ref{fig:prior_samples}.

Test data was synthesized by Fourier transforming the reference reconstruction images and corrupting with additive Gaussian noise using $\sigma = 0.01 \cdot \mathrm{max}\left(x_1^{GT}\right)$, and retrospectively under-sampling their Fourier coefficients with acceleration levels of $R=3,4,5$ using uniform random vertical sampling patterns with fully sampled autocalibration signal (ACS) regions of k-space (examples shown in Figure \ref{fig:cases}). Although this noise level is low with respect to $x_1$, $x_2$ has lower signal intensity leading to relatively lower SNR in simulated $x_2$ measurements. We note here that this is an inverse crime since test data was synthesized from single-coil images reconstructed from a $3$D multi-coil acquisition \cite{efrat_pnas}.

In the joint reconstruction scenario, we sampled both $x_1$ and $x_2$ at the prescribed acceleration level by collecting the same $12$ ACS lines and independent uniform random sampling for the remaining phase encode (PE) lines in each image contrast (Fig. \ref{fig:cases}). For the prior image conditioned reconstruction we only took under-sampled measurements from the target contrast $x_2$ by, again, collecting an ACS region and uniform random sampling for the remaining PE lines (Fig. \ref{fig:cases}). The test set included $100$ slices from three volumes ($300$ slices in total) not seen in the training or validation datasets. For each test image we conducted five different reconstructions: \textbf{(1)} Zero-filled IFFT (ZF), \textbf{(2)} Sample $x_2$ given only the fully sampled prior image $x_1$ ($E[x_2|x_1]$), \textbf{(3)} Sample $x_2$ using only its own under-sampled measurements $y_2$ ($E[x_2|y_2]$), \textbf{(4)} Sample $x_2$ from under-sampled measurements of both contrasts ($E[x_1,x_2|y_1,y_2]$), \textbf{(5)} Sample $x_2$ from under-sampled measurements $y_2$ and fully sampled prior image $x_1$ ($E[x_2|y_2, x_1]$). In each case we averaged $10$ posterior samples for the reconstruction. The step size $\eta_t$ was tuned individually for each method using a single validation patient with $100$ samples at $R=4$. 

To compare with end-to-end deep learning methods for joint reconstruction we trained an unrolled network based on MoDL \cite{modl} which shared a $4$ channel denoising network between the two contrasts (MC-MoDL). We trained the network for the joint reconstruction task by retrospectively under-sampling both contrasts at an acceleration level of $R=3$ as described above. This technique serves as a point of comparison to other unrolled end-to-end methods \cite{jVN}. 

After all the models had been trained we conducted three different experiments: \textbf{(1)} joint reconstruction at various acceleration levels, \textbf{(2)} joint reconstruction with altered forward operator (horizontal sampling mask), \textbf{(3)} and prior image guided reconstruction.

\subsection{Quantitative Metrics}
To evaluate reconstruction quality we used SSIM as defined in \cite{wang2004image} (higher is better) and normalized root mean squared error (NRMSE) between ground truth and reconstructed image pairs (lower is better), where NRMSE is defined between a ground truth image $x^{GT}$ and a reconstructed image $x^*$ as 

\begin{equation}
\label{eqn:nrmse}
NRMSE = \frac{\norm{x^{GT}-x^*}_2}{\norm{x^{GT}}_2}.
\end{equation}
Results in tables report average loss values across the entire test set for each metric. 

% \begin{figure*}[h!]
% \begin{center}
%   \includegraphics[width=16.5cm]{figures/single_score.png}
% \end{center}
% \caption{Knee samples drawn from a marginal score model $p(x_2)$ (Contrast 2).}
% \label{fig:single_score}
% \end{figure*}

% \begin{figure*}[h!]
% \begin{center}
%  \includegraphics[width=16.5cm]{figures/joint_score.png}
% \end{center}
% \caption{Knee samples from a joint score model trained on paired images. The $1^{st}$ and $3^{rd}$ columns are $x_1$ (Contrast 1) samples drawn from $p(x_1,x_2)$ with corresponding $x_2$ (Contrast 2) samples in columns $2$ and $4$. }
% \label{fig:joint_score}
% \end{figure*}

\section{Results}
\begin{figure*}[h!]
\begin{center}
  \includegraphics[width=18cm]{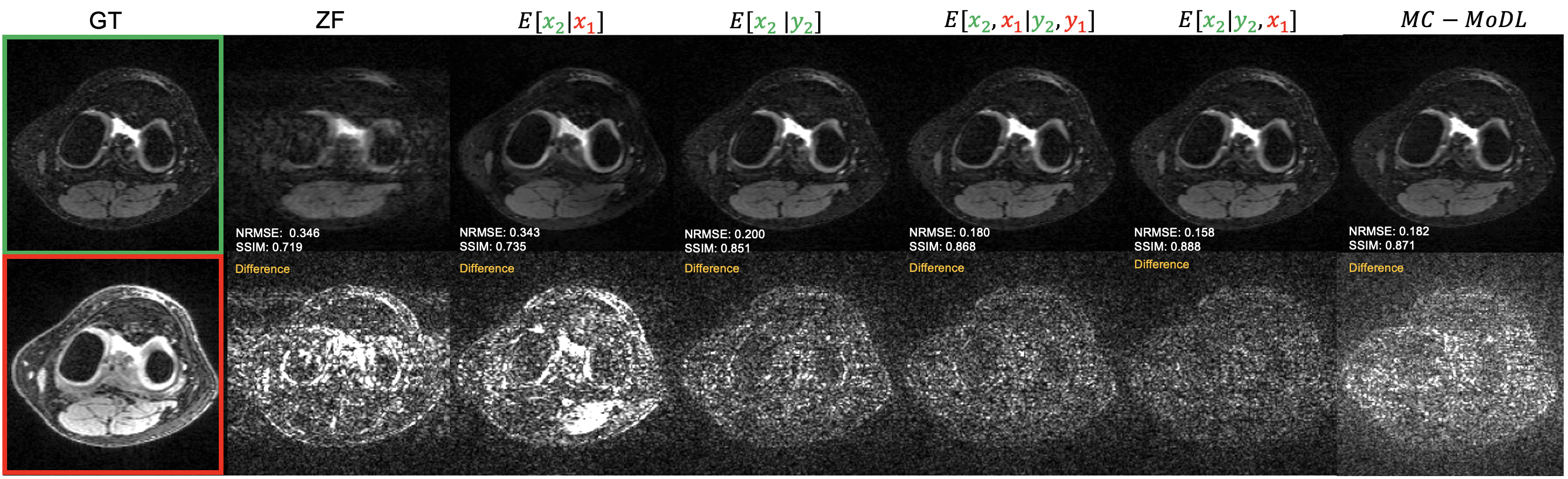}
\end{center}
\caption{Reconstruction of a slice from the test set at $R=3$ with vertical sampling. Left-most column shows the ground-truth (GT) $x_2$ and $x_1$. Top: reconstruction of $x_2$ given different side information. Bottom: Difference image with ground-truth.}
\label{fig:ex1}
\end{figure*}

\begin{figure}[h!]
\begin{center}
  \includegraphics[width=\columnwidth]{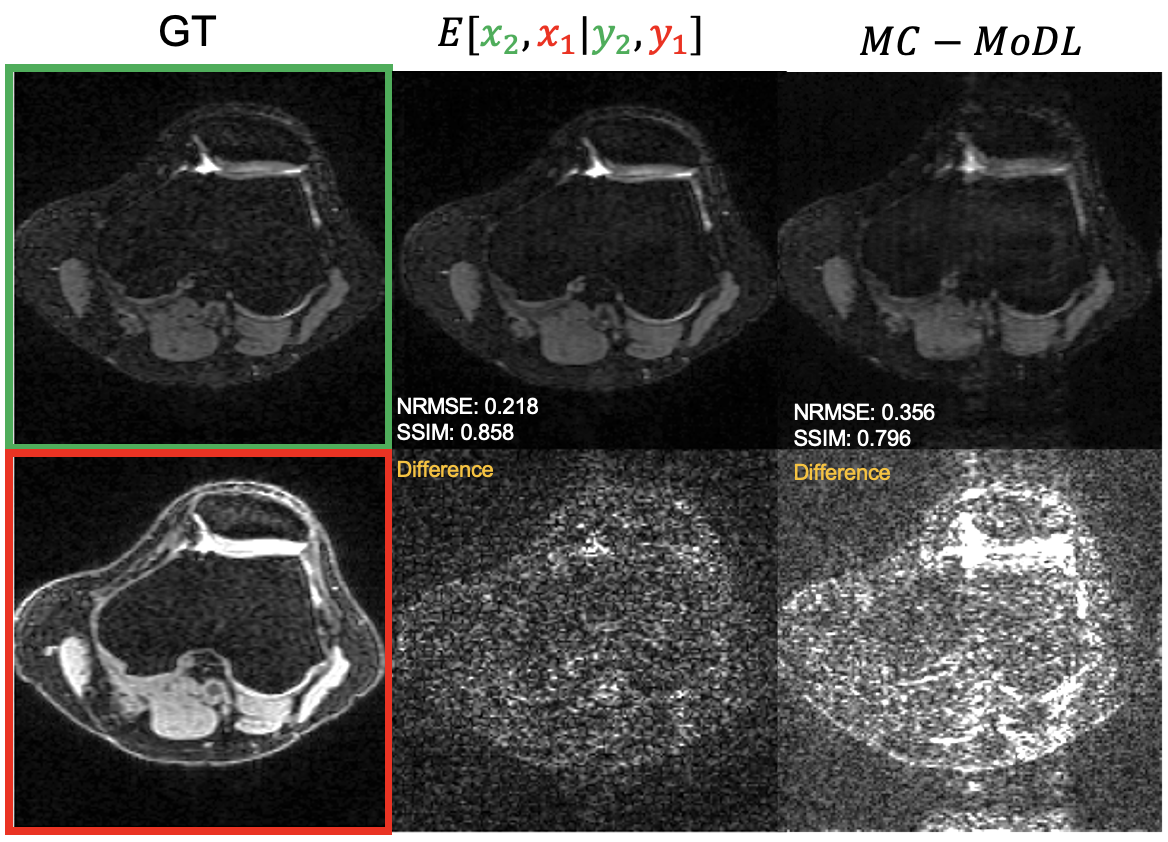}
\end{center}
\caption{Reconstruction of a slice from the test set using a horizontal sampling pattern at $R=3$.}
\label{fig:horiz_ex}
\end{figure}

Numerical results for reconstructing $x_2$ in the joint reconstruction case are shown in Table~\ref{table:table1} for each approach described above. Example reconstructions for $R=3$ are shown in Fig. \ref{fig:ex1}. The top and bottom images in the left most column are $x_2^{GT}$ and $x_1^{GT}$ respectively. The remaining columns each represent a different reconstruction technique as described in Section \ref{sec:exp}. For each reconstruction technique the top row is the reconstructed image and the bottom row contains the difference image between the reconstructed image and the ground truth image. All images are plotted on the same dynamic range and the difference images are scaled by a factor of $10$. Table~\ref{table:table2} shows the reconstruction metrics for when we apply the joint reconstruction techniques to data that has been sampled with a different pattern at $R=3$. Figure \ref{fig:horiz_ex} shows an example reconstruction for this case. Results for the prior guided image reconstruction experiment where we have access to a fully sampled prior image $x_1^{GT}$ are shown in Table \ref{table:table3}.

\begin{table*}[t]
\centering
\caption{Vertical Sampling Reconstruction Metrics}
\label{table:table1}
\begin{tabularx}{\linewidth}{ c *{11}{C} }
\\
    \toprule
\textbf{R}    & \multicolumn{2}{c}{ZF}
            & \multicolumn{2}{c}{$E[x_2|y_2]$}
                    & \multicolumn{2}{c}{$E[x_2|x_1]$}
                            & \multicolumn{2}{c}{$E[x_1,x_2|y_1,y_2]$} 
                            & \multicolumn{2}{c}{MC-MoDL}
                                 \\
   &   \small{\textbf{NRMSE}$\downarrow$}  & 
   \small{\textbf{SSIM$\uparrow$}}  &
   \small{\textbf{NRMSE$\downarrow$}}  &
   \small{\textbf{SSIM$\uparrow$}} &
   \small{\textbf{NRMSE$\downarrow$}}  &   \small{\textbf{SSIM$\uparrow$}}  &   \small{\textbf{NRSME$\downarrow$}}  &   \small{\textbf{SSIM$\uparrow$}}  &   \small{\textbf{NRMSE$\downarrow$}}  &   \small{\textbf{SSIM$\uparrow$}} 
                                \\
    \midrule
3   &   $0.416$   &   $0.707$   & $0.230$  &   $0.841$  &   $0.429$  &   $0.720$  &   $0.205$  &   $0.862$  &  $0.205$ &  $0.872$ \\
4   &   $0.446$  &   $0.690$  &   $0.265$  &   $0.801$  &   $0.429$  &   $0.720$  &   $0.236$  &  $0.834$ &  $0.236$ &  $0.849$ \\
5   &   $0.463$  &   $0.681$  &   $0.295$  &   $0.784$  &   $0.429$  &   $0.720$  &   $0.265$  &  $0.811$ &  $0.265$ &  $0.829$ \\
    \bottomrule
\end{tabularx}
\end{table*}

\begin{table}[t]
\centering
\caption{Horizontal Sampling Reconstruction Metrics}
\label{table:table2}
\begin{tabularx}{\linewidth}{ c *{7}{C} }
\\
    \toprule
\textbf{R}    & \multicolumn{2}{c}{ZF}
                            & \multicolumn{2}{c}{$E[x_1,x_2|y_1,y_2]$}
                            & \multicolumn{2}{c}{MC-MoDL}\\
   &   \small{\textbf{NRMSE}$\downarrow$}  & 
   \small{\textbf{SSIM$\uparrow$}}  &
   \small{\textbf{NRMSE$\downarrow$}}  &
   \small{\textbf{SSIM$\uparrow$}} &
   \small{\textbf{NRMSE$\downarrow$}}  &   \small{\textbf{SSIM$\uparrow$}} 
                                \\
    \midrule
3   &   $0.465$   &   $0.638$ &  $0.205$  &   $0.856$ &  $0.304$ &  $0.796$  \\
    \bottomrule
\end{tabularx}
\end{table}

\begin{table}[t]
\centering
\caption{Prior Image Guided Reconstruction Metrics}
\label{table:table3}
\begin{tabularx}{\linewidth}{ c *{7}{C} }
\\
    \toprule
\textbf{R}    & \multicolumn{2}{c}{ZF}
                            & \multicolumn{2}{c}{$E[x_2|y_2,x_1]$}
                            & \multicolumn{2}{c}{MC-MoDL}\\
   &   \small{\textbf{NRMSE}$\downarrow$}  & 
   \small{\textbf{SSIM$\uparrow$}}  &
   \small{\textbf{NRMSE$\downarrow$}}  &
   \small{\textbf{SSIM$\uparrow$}} &
   \small{\textbf{NRMSE$\downarrow$}}  &   \small{\textbf{SSIM$\uparrow$}} 
                                \\
    \midrule
3   &   $0.416$   &   $0.707$  &  $0.184$  &   $0.878$ & $0.234$ &  $0.861$   \\
4   &   $0.446$   &   $0.690$  & $0.200$  &   $0.861$& $0.269$ &  $0.842$  \\
5   &   $0.463$   &   $0.681$ & $0.211$  &   $0.850$ &  $0.297$ &  $0.828$  
\\
    \bottomrule
\end{tabularx}
\end{table}

\section{Discussion}
From our results we observe in Table \ref{table:table1} that using joint priors ($E[x_1,x_2|y_1,y_2]$) via score-based generative modeling for multi-contrast image reconstruction improves reconstruction quality when compared to marginal priors ($E[x_2|y_2]$). In the same set of experiments we are able to match the performance of end-to-end techniques (MC-MoDL) in-distribution (Table \ref{table:table1}). Our technique vastly outperforms end-to-end techniques when a distribution shift occurs in the forward model as seen in Table \ref{table:table2}. This is because end-to-end techniques are trained with access to a limited distribution of forward operators at training time and thus are known to generalize poorly to alterations in the forward model at test time. This is also shown in Table \ref{table:table3} where we give MC-MoDL access to fully sampled ground truth reference images $x_1^{GT}$ and its performance dips sharply even compared to the case where $y_1$ is under-sampled by a factor of $3$. This, again, can be attributed to the poor generalization of end-to-end techniques in the presence of distribution shifts. Our technique does not see the forward operator at training time and thus does not experience the shift at test time. We emphasize that for both joint reconstruction ($E[x_1,x_2|y_1,y_2]$) and prior image guided reconstruction ($E[x_2|y_2,x_1]$) we need not retrain our generative prior. This robustness to forward model shift becomes increasingly important in multi-contrast imaging since the degrees of freedom with which clinicians can alter acquisition scan parameters (alter the forward model) increases rapidly and training an end-to-end model for each use case is impractical. 

It should be noted that our experiments were conducted on distributions which already contained inherent noise and thus our image reconstruction metrics were inherently low even for $R=3$. Additionally, for the specific dataset used it is not realistic to only under-sample one DESS echo. Our joint reconstruction method, as with other joint reconstruction techniques, requires the reconstruction of images after all image contrasts of interest are collected. This has the possibility to inhibit many clinical imaging sessions as most exams use intermittent image reconstructions throughout the exam to guide later scan protocol selections. This issue, however, could be mitigated by first starting with an unconditioned reconstruction and iteratively conducting heavier conditioned reconstructions as more scans are completed.

\section{Conclusion}
Multi-contrast imaging is ubiquitous in clinical MRI exams due to the large variety of tissue parameters and their differing sensitivities to pathology. We have shown that by leveraging shared statistical information between images via a learned Bayesian prior we can achieve marked improvements over state-of-the-art reconstruction techniques for single contrast reconstructions and even end-to-end joint reconstruction techniques. A clear extension can be made to longitudinal imaging scenarios where patients undergo scans separated by a period of time to track disease progress \cite{long_mri}. Our technique could be used to condition reconstruction at a present study on prior examinations. We emphasize that this work has the potential to be applied to settings where we wish to sample from $p(x|y,c)$ where $c$ is not necessarily another image but rather additional side information (age, pre-existing conditions, etc...).

% \clearpage
\bibliographystyle{IEEEbib}
\bibliography{refs}

\end{document}